\documentclass[10pt, a4paper]{article}

\usepackage{lrec-coling2024} 

\usepackage{multirow}
\usepackage{booktabs}
\usepackage{amsmath,amsfonts}
\usepackage{amssymb}
\newtheorem{assumption}{Assumption}

\newcommand\blfootnote[1]{%
    \begingroup
    \renewcommand\thefootnote{}\footnote{#1}%
    \addtocounter{footnote}{-1}%
    \endgroup
}

\title{Document Set Expansion with Positive-Unlabelled Learning Using Intractable Density Estimation}

\name{Haiyang Zhang$^{1*}$, Qiuyi Chen$^{1*}$, Yuanjie Zou$^{1}$\\
{\bf \large Yushan Pan$^{1}$, Jia Wang$^{1}$, Mark Stevenson$^{2}$}}

\address{ $^{1}$Xi'an Jiaotong Liverpool University\\
    $^{2}$University of Sheffield\\
    \{haiyang.zhang, yushan.pan, Jia.wang02\}@xjtlu.edu.cn\\
    \{qiuyi.chen2002, yuanjie.zou\}@student.xjtlu.edu.cn\\
    mark.stevenson@sheffield.ac.uk\\}

\abstract{
    The Document Set Expansion (DSE) task involves identifying relevant documents from large collections based on a limited set of example documents. Previous research has highlighted Positive and Unlabeled (PU) learning as a promising approach for this task. However, most PU methods rely on the unrealistic assumption of knowing the class prior for positive samples in the collection. To address this limitation, this paper introduces a novel PU learning framework that utilizes intractable density estimation models. Experiments conducted on PubMed and Covid datasets in a transductive setting showcase the effectiveness of the proposed method for DSE.
    Code is available from
    \url{https://github.com/Beautifuldog01/Document-set-expansion-puDE}.
    \\ \newline 
}

\begin{document}

\maketitleabstract
\blfootnote{* denotes equal contribution.}

\section{Introduction}
We focus on the scenario where a user has access to a (possibly small) set of
documents of interest and wishes to identify further such documents within a
large collections, a problem known as Document Set Expansion (DSE)
\cite{EACL2021_DSE,DocRanking_SIGIR18,ECIR2022_seed_shuai}. DSE is a common
information seeking problem, for example when searching scientific literature
for papers that are similar to a small set of relevant `seed' publications
\cite{SIGIR22_seedCollection}.
It can also occur in the maintenance of curated databases of scientific
literature where examples of relevant studies are readily available but there
may not be an explicit query \cite{chen2021litcovid}.

Query-by-document (QBD) is an approach to DSE which involves treating the set
of documents as an extended query used to rank the documents in the collection
\cite{QBD2022ECIR,DocRanking_SIGIR18,QBD2009}. A common QBD approach focuses on
constructing an accurate query from the seed documents, using methods such as
bag-of-word \cite{QBD2009} or Monte-Carlo (MC) sampling procedure
\cite{QBD_MC2022}.
However, such methods fail to capture the local or global connections between
words \cite{EACL2021_DSE}. More recent work fine-tuned a BERT re-ranker for the
QBD retrieval task \cite{QBD2022ECIR}, but requires a fully labelled dataset to
train the neural models. In addition, the majority of the QBD approaches only
work with a single seed document \cite{QBD2022ECIR,DocRanking_SIGIR18}.

\begin{figure}
    \centering
    \includegraphics[width=77mm,height=36mm]{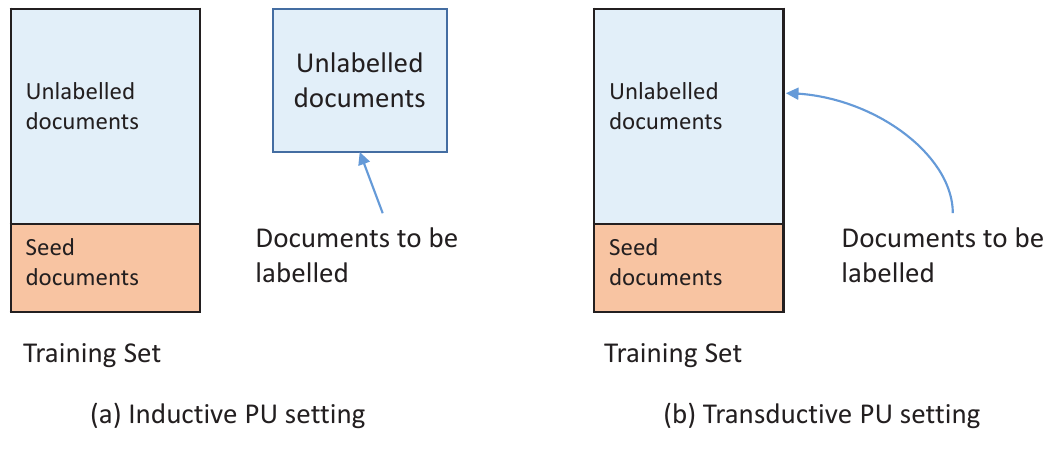}
    \caption{Inductive setting and transductive setting in PU learning.}
    \label{fig:setting difference}
\end{figure}

\newcite{EACL2021_DSE} treated the DSE task as a positive and unlabelled (PU) learning problem by learning a binary classifier using only positive and unlabelled data \cite{survey2020,uPUplessis15,nnPU2017}.
They introduce a new PU method based on Non-negative PU (nnPU) \cite{nnPU2017}, and show that their methods can outperform common information retrieval (IR) solutions for the DSE task.
However, some important issues remain unresolved:
\begin{itemize}
    \item PU methods that rely on misclassification risk, such as nnPU, assumes that the
          class prior, $\pi = P(Y=1)$, is known. The class prior denotes the proportion
          of positive data in the unlabelled data and plays an important role in PU
          learning.
          However, in practical applications, $\pi$ is usually unknown and it cannot be
          treated as a trainable parameter \cite{vpu2020}. Several studies propose to
          estimate the class prior as an intermediate step for PU classification
          \cite{prioresimation_2016ICML,ijcai2020PU_priorEstimation}. However, such
          methods commonly utilize complex kernel machines. Moreover, inaccurate
          estimation may bring more errors in the PU classification \cite{vpu2020}.
    \item DSE is essentially a transductive problem since we aim to identify all positive
          documents from the unlabelled set (U). In such a case, the unlabelled set
          should be used for both training and testing. However, \citet{EACL2021_DSE}
          treat DSE as an inductive problem, where U is split into training and test
          sets, with only samples in the test set being labelled. Such experimental
          settings cannot reflect the ground truth performance of the model for the DSE
          task. The difference between the two settings can be found in Figure
          \ref{fig:setting difference}.

\end{itemize}
To address these issues, we propose a novel PU learning framework based on intractable models, which
does not require a known class prior. It aims to learn a bayesian binary
classifier by merely making use of the distribution of labelled and unlabelled
data, without class prior involved in training. Intractable models, i.e. Kernel
Density Estimation (KDE) \cite{neural_ranker2022} and Energy-based model (EBM)
\cite{EBM2006} are used to estimate the density, as they do not restrict to the
tractability of the normalizing constant \cite{zhai2016EBM}.
Consequently, it does not require assumptions on the form of data distribution
to be fitted.
Experiments are conducted in a transductive setting to better reflect the DSE
task.

The contributions of this work are: 1) to identify the limitations of previous
for the DSE task \cite{EACL2021_DSE}; 2) propose \textbf{puDE}, a new PU
learning framework by using intractable models for density estimation that does
not require any knowledge of class prior
; 3) demonstrate that \textbf{puDE} outperforms state-of-the-art PU methods for the DSE task on real-world datasets.

\section{Background}

In the binary classification task, given $\mathbf{x} \in \mathcal{R}^{d}$ as
the input instance and $Y \in \{1, -1\}$ as the label of $\mathbf{x}$, the goal
is to learn a decision function $\Phi:{X}\rightarrow Y$ that can separate the
positive and negative examples. In order to facilitate the training of an
accurate classifier, it is assumed that the training data represents an
independent and identically distributed sample of the actual underlying
distribution: $ \mathbb{P}(\mathbf{x}) = \pi\mathbb{P}(\mathbf{x}|Y=1) +
    (1-\pi)\mathbb{P}(\mathbf{x}|Y=-1) $ where $\pi = P(Y=1)$ is the class prior.

The setting of PU is a special scenario within the binary classification
problem, where only a small portion of positive examples are observed
\cite{survey2020}. The training set is a combination of the labelled positive
set ${X}_{LP}$, the unlabelled positive set ${X}_{UP}$, and the unlabelled
negative set ${X}_{UN}$, such that ${X}={X}_{LP} \cup {X}_{UP} \cup {X}_{UN}$.
Let $s \in \{1,0\}$ present the label status of $y$ ($s = 1$ if labelled,
otherwise $s = 0$), there will be:
\begin{equation*}
    \begin{aligned}
        {X}_{LP} & = \{\mathbf{x}|s=1, Y=1\},\quad
        {X}_{UP} = \{\mathbf{x}|s=0, Y=1\}           \\
        {X}_{UN} & = \{\mathbf{x}|s=0, Y=-1\}, \quad
        {X}_{U}= \{\mathbf{x}|s=0\}
    \end{aligned}
\end{equation*}
The label frequency can be represented as  $c=P(s=1|Y=1)$ \cite{PU2008Eklan}.
In PU scenario, ${X}_{LP}$ are selected from a completed set of positive
examples ${X}_{P}$ under certain probabilistic labeling mechanism, and the
probability of an example being labelled is defined as
$e(\mathbf{x})=P(s=1|\mathbf{x}, Y=1)$, known as propensity score
\cite{PU2008Eklan}. Hence, the distribution of labelled positives $f_{LP}
    \triangleq P(\mathbf{x}|s=1, Y=1)$ can be seen as a biased version of $f_{P}
    \triangleq P(\mathbf{x}|Y=1)$: \ 
\begin{equation}
    \ 
    f_{LP}(\mathbf{x}) = \frac{e(\mathbf{x})}{c}f_{P}(\mathbf{x}),
\end{equation}
where $c=P(s=1|Y=1)$ is the label frequency. The goal of PU learn is to learn a binary classifier $\Phi:{X}\rightarrow Y$ that can separate the positive from unlabelled examples. In this work, our objective is to estimate an optimal Bayesian classifier under the following assumption:
\begin{assumption}
    The positive labelled data are randomly selected from the set of positive data and are  identically distributed with the positive unlabelled data: $f_{LP}(\mathbf{x}) = f_{P}(\mathbf{x})$, which is known as the \textit{Selected Completely At Random (SCAR)} assumption \cite{survey2020}.
\end{assumption}

\section{PU Learning with Intractable Models} \label{PUL}

We consider the following task: we have a set of labelled positive documents
${X}_{LP}$ on a fine-grained topic and want to find more documents about that
topic from a large unlabelled collection ${X}_{U}$. Given ${X}_{LP}$ and
${X}_{U}$, the objective of our method is to learn a Bayesian classifier $\Phi$
to approximate $\mathbb{P}(Y=1|\mathbf{x})$. According to the Bayesian rule, we
have:
\begin{equation}\small
    \mathbb{P}(Y=1|\mathbf{x})=\frac{\mathbb{P}(\mathbf{x}|Y=1)\mathbb{P}(Y=1)}{\mathbb{P}(\mathbf{x})}
    = \frac{f_p(\mathbf{x})}{f(\mathbf{x})}\pi
    \label{bayesian}
\end{equation}
where $f_p(\mathbf{x})$ is the positive data distribution and $f(\mathbf{x})$ is the distribution of the whole dataset. It is intuitive to estimate the probability density of $f_p(\mathbf{x})$ and $f(\mathbf{x})$ respectively, so that $\pi$ can be treated as a constant for each $\mathbf{x}$ and not involved in the training process. In such a way, we can learn a classifier without the need for class prior estimation which is an intermediate step for the PU classification task \cite{ijcai2020PU_priorEstimation}.

Let $p_\theta(\mathbf{x}):\mathcal{R}^d \rightarrow [0, 1]$ and
$q_\theta(\mathbf{x}):\mathcal{R}^d \rightarrow [0, 1]$ be the two models to
estimate $f_p(\mathbf{x})$ and $f(\mathbf{x})$, $\mathbb{P}(Y=1|\mathbf{x})$
can be then approximated by:
\begin{equation}
    \mathbb{P}(Y=1|\mathbf{x}) \approx \Phi(\mathbf{x}) = \frac{p_\theta(\mathbf{x})}{q_\theta(\mathbf{x})}\pi
    \label{biCls}
\end{equation}
Under Assumption 1, i.e. $f_{LP}(\mathbf{x}) = f_{P}(\mathbf{x})$, we can estimate $f_{P}$ using samples from $X_{LP}$. In this paper, we try to make less restriction on the underlying distribution on the data we fit. Therefore, intractable density estimation methods in both nonparametric and parametric forms are adopted.


\subsection{Nonparametric Density Estimation}
Kernel Density Estimation (KDE) is a nonparametric density estimation
technique, which has been applied in recommender systems and information
retrieval \cite{Silverman_2018,chakraborty2022kernelIR}.
For a given dataset $\{\mathbf{x}_1,\mathbf{x}_2,\cdots \mathbf{x}_n\}$, the
estimated density $\hat{f}$ at $\mathbf{x}$ using KDE is defined as:
\begin{equation}\small
    \widehat{f}_{kde}(\mathbf{x})= \frac{1}{n h} \sum_{i=1}^n K\left(\frac{\mathbf{x}-\mathbf{x}_i}{h}\right)
\end{equation}
where $h$ is the bandwidth hyperparameter, and $K$ is a non-negative kernel function.
In the DSE task, given a set of labelled documents $X_{LP}$, $\Phi(\mathbf{x})$
estimated with KDE is represented as follows:
\begin{equation}\small
    \Phi(\mathbf{x})= \frac{p_\theta(\mathbf{x})}{q_\theta(\mathbf{x})}\pi =\frac{\widehat{f}_{p,kde}(\mathbf{x})}{\widehat{f}_{kde}(\mathbf{x})}\pi\\
\end{equation}
where $\widehat{f}_{p,kde}$ is the estimated density of positive data which can be estimated by samples from ${X}_{LP}$, and $\widehat{f}_{kde}$ is the estimated density of the whole data ${X}$. Gaussian density function $K(\mathbf{x})=\frac{1}{\sqrt{2 \pi}} e^{-\frac{1}{2} \mathbf{x}^2}$ is used as the kernel function.

\subsection{Parametric Density Estimation}
The parametric approach used to estimate the density is the energy-based model
(EBM) \cite{EBM2006}, which is a powerful tool for representing complex
high-dimensional data distributions. It aims to learn an energy function that
assigns a low energy value to observed data and a high energy value to
different values. Compared with other parametric density estimation methods,
such as VAE \cite{kingma2013auto} and Masked Autoregressive Density Estimators
(MADE) \cite{papamakarios2017masked}, EBM does not make any assumption on the
form of data distribution they fit. An EBM parameterizes any probability
density for $\mathbf{x}\in \mathbb{R}^{d}$ as:
\begin{equation}
    \begin{aligned}
        f_{EBM,\theta}(\mathbf{x})=\frac{e^{-E_{\theta}(\mathbf{x})}}{Z_{{\theta}}}
        \quad \quad
        Z_{\theta}=\int e^{{-E_{\theta}}(\mathbf{x})}dx
    \end{aligned}
\end{equation}
where $E_{\theta}(\mathbf{x})$ is the energy function, which is a nonlinear regression function configured with optimal $\theta$, and $Z_{\theta}$ is the partition function, which is a function of $\theta$ but is a constant with respect to $\mathbf{x}$. For the DSE task, we use two neural networks ($g_{p_{\theta}}$ and $g_{q_{\theta}}$) as the energy function to estimate $p_{\theta}$ and $q_{\theta}$ respectively.
Thus, $\Phi(\mathbf{x})$  is rewritten as:
\begin{equation}
    \small
    \begin{aligned}
        \Phi(\mathbf{x}) & = \frac{p_\theta(\mathbf{x})}{q_\theta(\mathbf{x})}\pi
        =\frac{e^{-g_{p_{\theta}}(\mathbf{x})}}{Z_{p_{\theta}}} / \frac{e^{-g_{q_{\theta}}(\mathbf{x})}}{Z_{q_{\theta}}}\pi                  \\
                         & =e^{(g_{q_{\theta}}(\mathbf{x})-g_{p_{\theta}}(\mathbf{x}))}\left(\frac{Z_{q_{\theta}}}{Z_{p_{\theta}}}\pi\right)
        \label{PDE}
    \end{aligned}
\end{equation}
where $\frac{Z_{q_{\theta}}}{Z_{p_{\theta}}}\pi$ is a constant for each
$\mathbf{x}$ and can be ignored in practise. Hence, $\Phi(\mathbf{x})$ can be
approximated by the exponent: $\Phi(\mathbf{x}) :=
    g_{q_{\theta}}(\mathbf{x})-g_{p_{\theta}}(\mathbf{x})$.
\paragraph{Model Training}
We employ the maximum likelihood training with Markov Chain Monte Carlo (MCMC)
sampling to train the energy models, such that there will be no need to
calculate the constant term $\frac{Z_{q_{\theta}}}{Z_{q_{\theta}}}\pi$ during
the training process. With maximum likelihood estimation (MLE), we can fit
$p_{\theta}$ to $f_{LP}(\mathbf{x})$ and $q_{\theta}$ to $f(\mathbf{x})$ by
maximizing the following expected log-likelihood:
\begin{equation*}\small
    \begin{aligned}
        \mathrm{E}_{X_{LP}}\left[\log \mathrm{p}_{\theta}(\mathbf{x})\right] \quad\quad
        \mathrm{E}_{X}\left[\log \mathrm{q}_{\theta}(\mathbf{x})\right] \\
    \end{aligned}
\end{equation*}
which are equivalent to minimizing the following KL divergence:
\begin{equation*} \small
    \label{bayes_opt}
    \begin{aligned}
        \arg \min _\theta \mathrm{KL}(f_{LP}(\mathbf{x}) \| \mathrm{p_{\theta}}) \quad
        \arg \min _\theta \mathrm{KL}(f(\mathbf{x}) \| \mathrm{q_{\theta}}) \\
    \end{aligned}
\end{equation*}
where $ f_{LP}(\mathbf{x}) = f_{P}(\mathbf{x}) $, and $f(\mathbf{x})$ is the real distribution of positive data and the whole data, respectively.
The loss function to minimize is defined as:
\begin{equation}
    \begin{aligned}
        \alpha\left(-\mathrm{E}_{X_{LP}}\left[\log \mathrm{p}_{\theta}(\mathbf{x})\right]\right)
         & +\beta \left(-\mathrm{E}_{X}\left[\log \mathrm{q}_{\theta}(\mathbf{x})\right]\right) \\
    \end{aligned}
\end{equation}
where $\alpha$ and $\beta$ are coefficients.
By using the MCMC sampling approach, the gradient of the log-likelihood of
$p_{\theta}(\mathbf{x})$ and $q_{\theta}(\mathbf{x})$ are defined as:
\begin{equation*}\small
    \begin{aligned}
        \nabla_\theta \log p_\theta(\mathbf{x})=-\nabla_\theta g_{p_{\theta}}(\mathbf{x})-\mathrm{E}_{\mathbf{x} \sim p_\theta(\mathbf{x})}\left[-\nabla_\theta g_{p_{\theta}}(\mathbf{x})\right] \\
        \nabla_\theta \log q_\theta(\mathbf{x})=-\nabla_\theta g_{q_{\theta}}(\mathbf{x})-\mathrm{E}_{\mathbf{x} \sim q_\theta(\mathbf{x})}\left[-\nabla_\theta g_{q_{\theta}}(\mathbf{x})\right]
    \end{aligned}
    \label{gradLab}
\end{equation*}
The first terms in both equations above are straightforward to obtain. To approximating the second terms,  Langevin Dynamics is used to sample from $p_{\theta}(\mathbf{x})$ and $q_{\theta}(\mathbf{x})$:
\begin{equation*}\small
    \begin{aligned}
        \mathbf{x}_{t+1}=\mathbf{x}_t+\frac{\varepsilon}{2} \nabla \log p_{\theta}\left(\mathbf{x}_t\right)+\mathcal{N}(0, \varepsilon) \\
        \mathbf{x}_{t+1}=\mathbf{x}_t+\frac{\varepsilon}{2} \nabla \log q_{\theta}\left(\mathbf{x}_t\right)+\mathcal{N}(0, \varepsilon)
    \end{aligned}
    \label{sampleLab}
\end{equation*}
where $t$ denotes the iteration step, $\mathcal{N}$ is the normal distribution.
Since Langevin dynamics can be unreliable in high-intensity areas for
high-dimensional datasets, which will effect the model performance. To address
this issue, we add a risk estimator in the loss function:
\begin{equation}
    \begin{aligned}
        \alpha\left(-\mathrm{E}_{X_{LP}}\left[\log \mathrm{p}_{\theta}(\mathbf{x})\right]\right)
         & +\beta \left(-\mathrm{E}_{X}\left[\log \mathrm{q}_{\theta}(\mathbf{x})\right]\right) \\
         & +\gamma\left(R_{\ell_{0-1}}(\Phi(\mathbf{x}),s)\right)
    \end{aligned}
\end{equation}
where $\gamma$ is a coefficient and it decreases as training progresses, $R_{\ell_{0-1}}(\Phi(\mathbf{x}),s)$ represents the loss generated by binary classification using `s' as the label.
\section{Experiment}
\subsection{Dataset}
Experiments use PubMed datasets on three fine-grained topics derived from
\newcite{EACL2021_DSE}. Additionally, we use the Covid-19 dataset that is used
for Covid-19 literature classification \cite{shemilt2022machine} to simulate
real-world literature curation. All datasets were originally designed for
inductive classification, where each dataset is split into training,
validation, and testing sets. In our experiments, to simulate real-world DSE
(transductive case), we treat the test set in original data slipt settings as
${X}_{U}$ and use ${X}_{U}$ for both training and testing (${X}_{LP}$ and
${X}_{U}$ for training and ${X}_{U}$ for testing). Following
\newcite{EACL2021_DSE}, the number of labelled positives |LP| is set to
$\{20,50\}$ on Pubmed datasets. For the Covid-19 dataset, the labelled
positives are randomly sampled from their original positive training set, and
the number of $|LP|$ is set with respect to the ratio of ${X}_{LP}$ over
${X}_{U}$, ranging from 0.01 to 0.1 with step of 0.01, and from 0.1 to 1 with
step 0.1. The statistics of each set is summarized in Table 1.

\subsection{Comparison Methods}
The performance of puDE-\textit{kde} and puDE-\textit{em} are compared with the
following methods:
\begin{itemize}
    \item \textbf{BM25} BM25 \cite{robertson1995okapi} serves as a strong baseline in various IR tasks. In our paper, following the method in \newcite{EACL2021_DSE}, we vary the number of top documents (K) to be considered as positive examples, $K\in \{i\}^{5000}_{|LP|}$, and report the F1 mean and standard deviation across the $5000-|LP|$ cases.
    \item \textbf{nnPU} nnPU \cite{nnPU2017} is a recent PU method that are based on unbiased risk estimators. It is used as the baseline in various PU studies, and is the first method being used for DSE task \cite{EACL2021_DSE}.
    \item \textbf{VPU} VPU \cite{vpu2020} is the a state-of-the-art PU method that do not require knowledge of class prior. It uses a variational principle to modeling the error of the Bayesian classifier directly from the provided data.
\end{itemize}
PU classifiers, i.e. nnPU and VPU, were implement them in transductive fashion to complement the DSE task.


\subsection{Settings}
We use puDE-\textit{kde} and puDE-\textit{em} to denote our proposed PU models
that is based on KDE and EBM, respectively. For puDE-\textit{kde}, the
bandwidth is set to 1.9 for both $\widehat{f}_{p,kde}$ and $\widehat{f}_{kde}$,
and Gaussian function is used as the kernel. Since KDE suffers from the curse
of dimensionality, we use Variational Autoencoders (VAE)
\cite{ReductionVAE2020} with
50 latent dimensions to reduce the high text dimension in this work.  For puDE-\textit{em}, we use 512D 4-layer fully connected neural network as the energy function for $g_{p_{\theta}}$ and $g_{q_{\theta}}$.
The weights for the total loss function are set as $\alpha=1$, $\beta=1$, and
$\gamma=1$.

nnPU is implemented using the tricks from \newcite{EACL2021_DSE} but in
transductive version. For both nnPU and VPU, the classifiers are modeled by
512D 4-layer fully connected neural network, with Batch normalization
\cite{ioffe2015batch} and leaky ReLU \cite{maas2013rectifier} applied. Adam
optimizer with a learning rate of 1e-3 is employed. For all methods, SciBERT is
used as the pre-trained embedding. Following \newcite{EACL2021_DSE}, F1 score
is used as the evaluation metric.

\begin{table*}[htp]
    \small
    \centering
    \label{dataset1}
    \begin{tabular}{p{4.5cm}cp{0.8cm}p{0.8cm}p{0.8cm}}
        \hline
        \text{dataset}           & |LP| & $N_{U}$ & $N_{UP}$ & $N_{UN}$ \\
        \hline
        \multirow{2}{*}{Animals+Brain+Rats}
                                 & 20   & 10012   & 1844     & 8168     \\
                                 & 50   & 10027   & 2568     & 7459     \\\hline
        Adult+Middle Aged        & 20   & 10012   & 2881     & 7131     \\ +HIV infections & 50                       & 10027 &
        3001                     & 7026                                 \\\hline
        Renal Dialysis + Chronic & 20   & 7198    & 1201     & 5997     \\ Kidney Failure+ Middle
        Aged                     & 50   & 10025   & 1916     & 8109     \\\hline \text{Covid} & $\{47..4722\}$           & 4722 &
        2310                     & 2412                                 \\

        \hline
    \end{tabular}
    \caption{Statistics of ${X}_{U}$ for each set, where $N_{U}$, $N_{UP}$ and $N_{UN}$, the total number of unlabelled samples, the number of true positive samples and true negatives in the training set.
    }
\end{table*}
\begin{table*}[ht]

    \centering
    \begin{tabular}{p{4.5cm}p{0.8cm}cp{1.3cm}p{1.3cm}p{1.3cm}p{1.3cm}}
        \toprule
        Topic                                        & |LP|           & BM25                          & nnPU  & VPU   & puDE-\textit{kde} & puDE-\textit{em} \\
        \midrule \multirow{2}{*}{Animals+Brain+Rats} & 20             & $32.25 \pm 11.6$              & 33.03 &
        25.62                                        & 37.31          & \textbf{40.59}                                                                       \\       & 50                                 & $32.80 \pm 10.9$ & 38.76 & 29.32 &
        44.65                                        & \textbf{44.91}                                                                                        \\\hline
        Adult+Middle Aged                            & 20             & $26.75\pm 7.22$               & 31.30 & 29.77 & 36.18             &
        \textbf{39.67}                                                                                                                                       \\ +HIV infections & 50                                 & $31.85\pm 10.7$ & 34.16 & 31.42 &
        44.03                                        & \textbf{46.22}                                                                                        \\\hline
        Renal Dialysis + Chronic                     & 20             & $\boldsymbol{41.23 \pm 8.95}$ & 27.76 & 21.59 &
        36.63                                        & 35.59                                                                                                 \\ Kidney Failure+ Middle Aged & 50             & $35.78\pm 9.13$           & 32.84 &
        19.42                                        & \textbf{36.63} & 36.57                                                                                \\ \bottomrule
    \end{tabular}
    \caption{F1 comparison against baseline and state-of-the-art DES methods with transductive setting.}
    \label{tab:f1_comparison}
\end{table*}

\subsection{Results}

The F1 results across all methods are reported in Table
\ref{tab:f1_comparison}, where the best performance are shown in bold font.
Performance of nnPU is much worse than that reported by \newcite{EACL2021_DSE}
and is similar to BM25, which indicate that the PU solutions proposed in
\cite{EACL2021_DSE} is not as effective as they stated for the DSE task in
transductive setting. Both puDE methods outperform other methods, with one
exception where BM25 get the best result on the last topic. It should be
noticed that result reported for BM25 is the average across 5000-|LP| cases,
which is not the direct classification result and it serves as references to
the state-of-the-art \cite{EACL2021_DSE}. Both puDE methods show significant
improvement over nnPU and VPU, demonstrating that the proposed PU framework
based on density estimation is a better alternative for the DSE task.

Figure 1 demonstrates the F1 score for all methods on Covid-19 dataset, with
the ratio of |LP| over |U| ranging from 0.01 to 1. It can be seen that nnpu and
VPU get stable results only when more than 20\% of labelled data is available.
Both puDE methods perform well with less data (<10\%) and consistently shown
significant improvements over other methods with the increase of labelled data.
As a state-of-the-art PU method, VPU can get similar performance with our
models when the label ratio goes over 0.5. However, when the the number of
labelled positives is small, the performance is poor. This is due to its
training strategy, where equal batch size of unlabelled (U) and labelled (LP)
samples are fed into the model to calculate the variational loss. When $|LP|
    \ll |U|$, the distribution of LP may be different from that of U. Replicating
LP, until the size of LP equals the size of U, can lead to instability in the
model, making it difficult for the model to converge during training and
resulting in poor prediction performance.
\begin{figure}[tp]
    \centering
    \includegraphics[width=75mm,height=45mm]{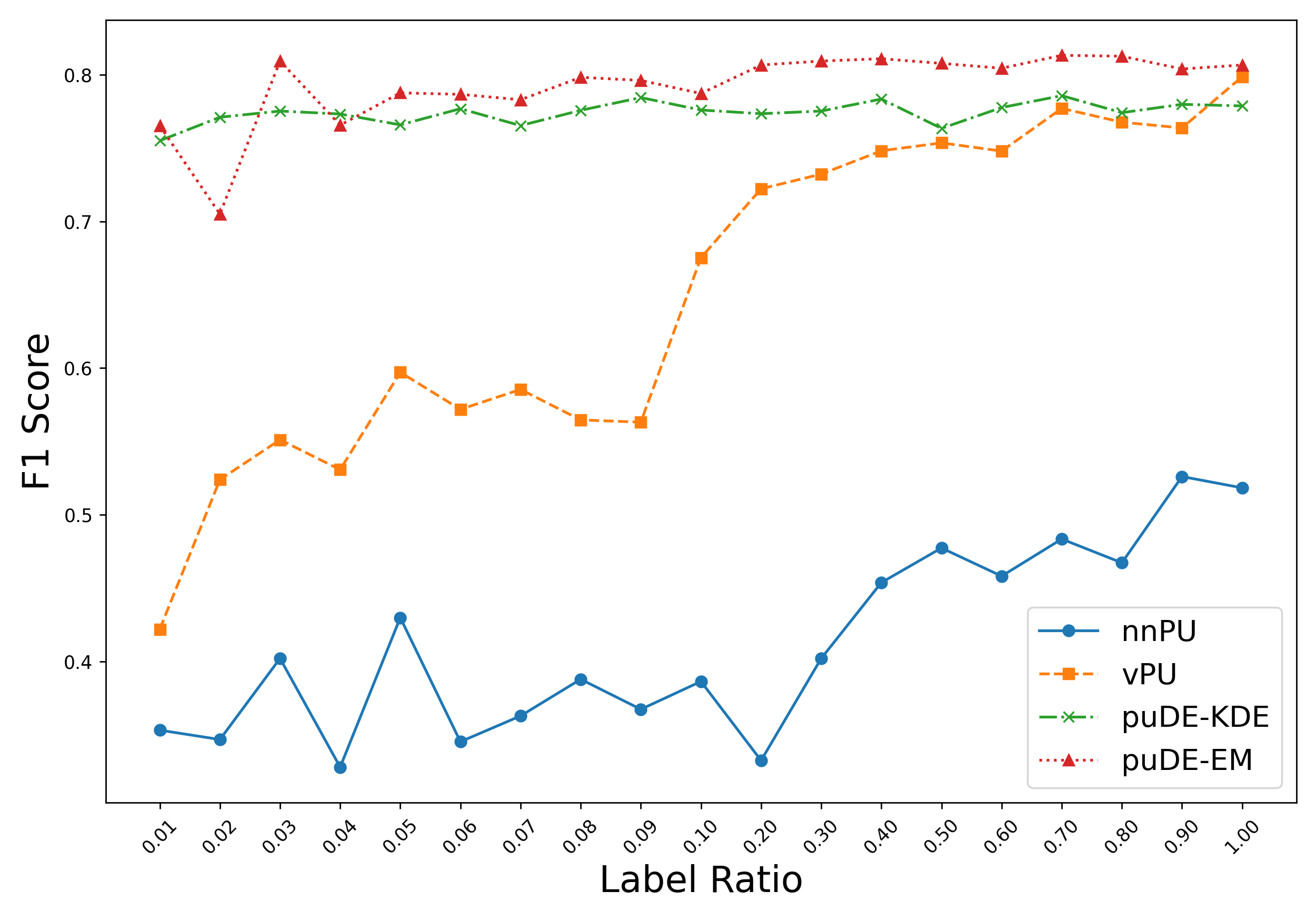}
    \caption{F1 comparison on covid dataset with respect to the ratio of |LP| over |U| ranging from 0.01 to 0.1 with step of 0.01 and from 0.1 to 1 with step of 0.1.}
    \label{AUC}
    \vspace{-2mm}
\end{figure}

We further conduct experiments for ranking task on Covid dataset to simulate
screening process in literature curation. The experiment settings are the same
as previous ones except that ranking-based evaluation metrics for systematic
reviews \cite{kanoulas2019clef,neural_ranker2022} are adopted. Precision at top
k\% documents in U (p@k\%), and recall at top k\% documents in U (r@k\%) are
reported. Table 3 shows the ranking effectiveness of all methods with the
number of labelled documents equals 50. It can be seen that our methods
produces the best overall performance.

\begin{table}[t]
    \label{tab:ranking}
    \centering
    \begin{tabular}{cp{1cm}p{1cm}p{1cm}p{1cm}}
        \toprule
        method         & P@10\%         & P@20\% & R@10\% & R@20\% \\ \midrule

        BM25           & 54.66          & 52.64  & 11.16  & 21.51  \\ nnPU & 52.54& 67.16& 10.74&27.45 \\
        VPU            & 56.77          & 57.30  & 11.90  & 23.41  \\ puDE-\textit{kde} & 70.26                     & 72.88 &
        \textbf{16.91} & 28.67                                     \\ puDE-\textit{em} & \textbf{76.91}                     & \textbf{75.11} &
        15.71          & \textbf{30.69}                            \\ \bottomrule
    \end{tabular}
    \caption{Performance comparison for ranking task on Covid dataset with |LP| = 50.}
\end{table}

\section{Conclusion}
This paper addresses the limitations of previous Positive-Unlabeled (PU)-based
approaches in solving the Document Set Expansion (DSE) task
\cite{EACL2021_DSE}. It demonstrates that experimental results obtained from an
inductive setting cannot be directly transferred to a real-world transductive
DSE scenario. To overcome these challenges, we propose a novel PU learning
framework based on intractable density estimation methods. A key advantage of
our approach is that it does not rely on prior knowledge of class proportions.
Experimental results validate the effectiveness of our proposed methods. In
conclusion, we assert that our approach represents a superior solution for the
DSE task compared to existing methods.

\section*{Acknowledgements}

We would like to express our gratitude for the support provided by the XJTLU AI
University Research Centre and the Jiangsu Province Engineering Research Centre
of Data Science and Cognitive Computation at XJTLU. Additionally, we
acknowledge the support from the SIP AI Innovation Platform (YZCXPT2022103) and
the Research Development Funding (RDF) at Xi'an Jiaotong-Liverpool University,
under contract numbers RDF-21-02-044 and RDF-21-02-008.

\section*{References}
\vspace{-5eX}

\bibliographystyle{lrec-coling2024-natbib.bst}
\bibliography{lrec-coling2024-example.bib}


\end{document}